\documentclass[pdflatex,sn-apa]{sn-jnl}

\usepackage{graphicx}%
\usepackage{multirow}%
\usepackage{amsmath,amssymb,amsfonts}%
\usepackage{amsthm}%
\usepackage{mathrsfs}%
\usepackage[title]{appendix}%
\usepackage{xcolor}%
\usepackage{textcomp}%
\usepackage{manyfoot}%
\usepackage{booktabs}%
\usepackage{algorithm}%
\usepackage{algorithmicx}%
\usepackage{algpseudocode}%
\usepackage{listings}%
\usepackage{enumitem}
\usepackage{url}
\usepackage{mathrsfs}

\theoremstyle{thmstyleone}%
\theoremstyle{thmstyletwo}%

\theoremstyle{thmstylethree}%

\begin{document}

\title[Article Title]{A real-time metric of online engagement monitoring 
}


\author*[1]{\fnm{Laura J.} \sur{Johnston}}\email{laura.johnston.22@ucl.ac.uk}

\author[1]{\fnm{Jim E.} \sur{Griffin}}\email{j.griffin@ucl.ac.uk}

\author[1]{\fnm{Ioanna} \sur{Manolopoulou}}\email{i.manolopoulou@ucl.ac.uk}

\author[1]{\fnm{Takoua} \sur{Jendoubi}}\email{t.jendoubi@ucl.ac.uk}

\affil[1]{\orgdiv{Department of Statistical Science}, \orgname{University College London}, \city{London}, \country{England}}


\abstract{Measuring online behavioural student engagement often relies on simple count indicators or retrospective, predictive methods, which present challenges for real-time application. To address these limitations, we reconceptualise an existing course-wide engagement metric to create a chapter-based version that aligns with the weekly structure of online courses. Derived directly from virtual learning environment log data, the new metric allows for cumulative, real-time tracking of student activity without requiring outcome data or model training. We evaluate the approach across three undergraduate statistics modules over two academic years, comparing it to the course-wide formulation to assess how the reconceptualisation influences what is measured. Results indicate strong alignment from as early as week 3, along with comparable or improved predictive validity for final grades in structured, lecture-based contexts. By the course midpoint, the weekly metric identifies as many low-performing students as are identifiable by the end of the course. While performance varies across modules, the chapter-based formulation offers a scalable and interpretable method for early engagement monitoring and student support.
}




\maketitle



\section{Introduction}
\label{sec:intro}

Student engagement, defined as ``the time and effort students devote to activities'' \citep[p.~683]{kuhWhatStudentAffairs2009}, is widely recognised as a multidimensional construct encompassing behavioural, emotional, and cognitive aspects \citep{fredricksSchoolEngagementPotential2004}. It has been positively linked to a range of academic outcomes, including improved learning, higher performance, greater satisfaction, and retention \citep{fredricksStudentEngagementContext2016, borupAcademicCommunitiesEngagement2020, leeRelationshipStudentEngagement2014, luOnlineLearningMeanings2020, xiongExaminingRelationsStudent2015}. Given its importance, there is increasing demand for scalable methods to measure and monitor engagement, particularly in online learning environments, which now account for a substantial proportion of learning in higher education. Capturing online engagement can help identify students in need of support and inform course and instructional design \citep{caspari-sadeghiApplyingLearningAnalytics2022}.

Measuring engagement in online contexts presents unique challenges. Traditional methods such as classroom observation, interviews, and self-report surveys are difficult to scale and often unreliable in digital settings \citep{fredricksMeasurementStudentEngagement2012, kahuFramingStudentEngagement2013}. In response, researchers have increasingly turned to log data from virtual learning environments (VLEs), such as Moodle, which offer unobtrusive, scalable records of students’ interactions with course content. These digital traces have been widely used to approximate the behavioural dimension of engagement \citep{henrieMeasuringStudentEngagement2015}, defined as students’ sustained participation in learning activities and observable effort in response to instructional tasks and contexts \citep{fredricksSchoolEngagementPotential2004, nkomoSynthesisStudentEngagement2021}. This study focuses on developing a real-time metric that captures part of online behavioural engagement in digital learning environments, using VLE log data to track students’ observable interactions with course resources.

Recent work has explored multiple approaches for monitoring online behavioural engagement using VLE log data. One example of proxy quantification of engagement is the metric proposed by \citet{chongProxyVariablesOnline2019}, which integrates five VLE indicators into a single composite score. Grounded in engagement theory, this metric defines online behavioural engagement as sustained, observable interaction with learning activities and validates the metric through its association with academic performance. However, it is designed to operate retrospectively, requiring complete course data and aggregating indicators across the entire course duration, which limits its utility for real-time use \citep{conijnPredictingStudentPerformance2017, zhangInvestigatingReliabilityAggregate2024}.

This example highlights a broader challenge in measuring behavioural engagement: although VLE log data can reflect student activity, it remains difficult to determine whether any given metric meaningfully captures the underlying construct \citep{henrieMeasuringStudentEngagement2015}. \citet{winneConstructConsequentialValidity2020} emphasises the importance of grounding such measures in established theory to ensure conceptual alignment. At the same time, empirical validation offers a practical and commonly used strategy. The well-documented relationship between engagement and academic performance provides a basis for evaluating metrics, with meaningful measures expected to show positive associations with academic outcomes  \citep{finchamCountingClicksNot2019}. 

Other approaches focus on real-time engagement monitoring for early warning systems. Learning analytics tools like the Moodle Engagement Analytics Plugin (MEAP) and MEAP+ use individual indicators, such as logins, submissions, and forum posts, to label students as “at risk” based on instructor-defined thresholds \citep{liuEnhancedLearningAnalytics2015, liuValidatingEffectivenessMoodle2015, nkomoSynthesisStudentEngagement2021}. Alternatively, a growing body of research employs supervised machine learning to predict academic outcomes from log-derived behavioural features \citep{brdnikUtilizingInteractionMetrics2022}. These models can be adapted to work with partial data, allowing engagement to be inferred earlier in the course \citep{okuboNeuralNetworkApproach2017}. 

Despite promising performance in retrospective analyses, both approaches present significant limitations when applied in real-time contexts. Tools like MEAP are simple to implement, but rely on instructor-defined thresholds that are often arbitrary and lack theoretical justification. Although \citet{liuValidatingEffectivenessMoodle2015} found associations between MEAP risk ratings and academic performance, the thresholds were determined retrospectively, limiting their applicability in live teaching settings where such cut-offs are unknown in advance. Predictive modelling approaches, meanwhile, are typically trained on historical engagement and grade data. Even when constrained to early-course subsets of data, these models still require outcome data during training and must generalise to new cohorts or courses if used for real-time monitoring, which is rarely achievable \citep{fazilNovelDeepLearning2024, dingTransferLearningUsing2019, boyerTransferLearningPredictive2015}. These limitations point to the need for an alternative approach that can provide engagement tracking without relying on retrospective tuning or training.

This study proposes a new metric for online behavioural engagement that directly addresses the limitations of existing approaches. Building on both the theoretical foundation and structural design of the \citet{chongProxyVariablesOnline2019} score, we reconceptualise their retrospective course-wide metric into a weekly, chapter-aligned format that enables real-time engagement monitoring. Unlike threshold-based tools or outcome-trained models, our approach depends solely on VLE interaction data and can be applied at any point in the course. We evaluate the metric by examining whether it captures similar behavioural signals to the course-wide formulation by \citet{chongProxyVariablesOnline2019}, how early in the course it begins to predict final grades, and its potential to support the early identification of at-risk students.

Section~\ref{sec:lit} reviews literature regarding the definition and measurement of online engagement. Section~\ref{sec:meth} introduces the dataset, describes the construction of our weekly engagement metric, and outlines the analytical methods employed. Section~\ref{sec:results} presents our findings, while Section~\ref{sec:discussion} discusses their practical implications and limitations.


\section{Related work}
\label{sec:lit}

\subsection{Defining online behavioural student engagement}

The definition of student engagement remains debated due to its complexity and multidimensional nature \citep{azevedoDefiningMeasuringEngagement2015, taiBewareSimpleImpact2019}. \citet{fredricksSchoolEngagementPotential2004} proposed a foundational framework encompassing three key dimensions: behavioural (observable participation, such as class attendance and completing tasks), emotional (students' affective reactions, like interest or boredom), and cognitive engagement (mental investment, including strategic planning and self-regulation). Subsequent research has further extended this framework to include dimensions such as agentic \citep{reeveAgencyFourthAspect2011}, social \citep{reschlyJingleJangleConceptual2012}, and academic engagement \citep{joksimovicHowWeModel2018}. These dimensions, although distinct, are interconnected and frequently influence each other. For instance, cognitive engagement (planning how to approach learning) might facilitate behavioural engagement (completing learning tasks) \citep{fredricksSchoolEngagementPotential2004}.

The initial framework of \citet{fredricksSchoolEngagementPotential2004} has been criticised for its overly broad inclusion of non-classroom activities. In response, \citet{wongStudentEngagementCurrent2022} proposed the Dual Component Framework, which differentiates between learning engagement (active involvement in classroom-based activities) and broader school engagement (extracurricular and community activities). They argue that behavioural learning engagement can be effectively assessed through observations of students' on-task behaviours and the time they actively dedicate to learning tasks.

The transition to online and blended learning environments in higher education has highlighted the need to revise existing definitions and measures of engagement for digital contexts.. \citet[p.~165]{martinOnlineLearnerEngagement2022} adapted the definition specifically for online settings, defining behavioural engagement as ``physical behaviours and energy that students demonstrate when completing learning activities [...] in the online learning environment''. \citet{henrieMeasuringStudentEngagement2015} support using VLE logs to capture behavioural engagement online, emphasising the importance of developing practical and validated measures using these digital traces.

In this paper, we focus specifically on measuring online behavioural engagement, aligning our definition with the learning engagement perspective proposed by \citet{wongStudentEngagementCurrent2022}, which emphasises active interaction with instructional activities. Extending this to digital settings, we adopt a definition consistent with \citet{henrieMeasuringStudentEngagement2015} and \citet{martinOnlineLearnerEngagement2022}, operationalising online behavioural engagement as observable student actions captured through VLE logs, such as accessing resources, completing activities, or participating in forums, that reflect active participation and responsiveness to online course content.

\subsection{Measuring online behavioural engagement}

A substantial body of research quantifies online behavioural engagement through proxy indicators derived from student activities in VLEs. These studies utilise behavioural signals reflecting interactions with digital resources, which are categorised into frequency-based features indicating the volume of student participation in online activities, and temporal features capturing the timing, pacing, and consistency of that participation.

Frequency-based indicators include total clicks, logins or sessions, time spent online, and resources accessed \citep{huDevelopingEarlyWarning2014, hoffmanMattersFrequencyImmediacy2023, motzValidityUtilityActivity2019, saqrLongitudinalTrajectoriesOnline2021a}. Some studies extend this by incorporating measures that count the number of different types of resources accessed, such as quizzes, videos, and forums. These indicators capture the variety of students' engagement behaviours alongside how much students engage \citep{youIdentifyingSignificantIndicators2016, motzValidityUtilityActivity2019}.

Temporal features capture the timing and consistency of engagement. Examples include promptness in accessing new materials \citep{bakerAnalyzingEarlyAtRisk2015, tripathiCognitiveEngagementScale2025} or the procrastination of submitting assignments \citep{cerezoStudentsLMSInteraction2016}, as well as the regularity or spacing of interactions \citep{conijnPredictingStudentPerformance2017, saqrLongitudinalTrajectoriesOnline2021a}. Temporal indicators are particularly relevant online, where students have greater control over pacing, or in flipped classrooms where success depends on engaging with materials before scheduled sessions \citep{jovanovicPredictivePowerRegularity2019}.

Studies often combine both indicator types to capture the complexity of behavioural engagement. \citet{conijnPredictingStudentPerformance2017} found that frequency-based indicators, such as the number of sessions and the diversity of resources accessed, were strong early predictors of student performance. However, frequency alone is often insufficient. \citet{hoffmanMattersFrequencyImmediacy2023} showed that immediacy and regularity added predictive value beyond frequency. Similarly, \citet{youIdentifyingSignificantIndicators2016} found that timely submission and consistent study patterns were more strongly associated with achievement than simple activity counts. These findings suggest that frequency and temporal indicators capture distinct but complementary aspects of online behavioural engagement. 

While many studies use these indicators directly as input features, others have sought to combine them into a single, interpretable metric. The metric proposed by \citet{chongProxyVariablesOnline2019}, which underpins this study, adopts precisely this approach, integrating both frequency and temporal indicators into a composite score designed to reflect online behavioural engagement. We describe the construction of this metric in detail in Sections~\ref{CW} and \ref{methods_RQ1}.

Beyond metric construction, behavioural indicators can be applied in a variety of ways. One common application is to explore student behaviour directly, for example, by clustering students based on interaction patterns. Unsupervised approaches, such as k-means and expectation-maximisation, group students into engagement profiles, which can then be used to monitor engagement trajectories across a course or degree programme, or to relate profiles to academic performance \citep{saqrLongitudinalTrajectoriesOnline2021a, cerezoStudentsLMSInteraction2016, heikkinenLongitudinalStudyInterplay2025}.

Alternatively, these indicators are often used as input features in predictive models. Supervised learning methods have been widely applied to forecast academic outcomes such as final grades, dropout risk, or instructor-rated engagement \citep{huDevelopingEarlyWarning2014, conijnPredictingStudentPerformance2017, motzValidityUtilityActivity2019}. Common modelling approaches include linear and logistic regression \citep{gasevicLearningAnalyticsShould2016, taylorLikelyStopPredicting2014, brdnikUtilizingInteractionMetrics2022}, as well as more complex machine learning algorithms, such as random forests and support vector machines \citep{umerPredictingAcademicPerformance2017, huDevelopingEarlyWarning2014, chukwuemekaEnhancedStudentEngagement2023, akcapinarUsingLearningAnalytics2019, badalPredictiveModellingAnalytics2023}. More recently, deep learning techniques, including recurrent neural networks, have been introduced to improve predictive performance \citep{okuboNeuralNetworkApproach2017, fazilNovelDeepLearning2024}. However, these models lack interpretability and demand substantial technical expertise to implement.

While predictive models often perform well when trained and evaluated on a single course, they frequently fail to generalise to new contexts such as different cohorts, courses, or platforms. Accuracy tends to decline when models are applied outside their original context due to differences in instructional design, types of resources, or student demographics \citep{boyerTransferLearningPredictive2015, dingTransferLearningUsing2019}. For instance, \citet{motzValidityUtilityActivity2019} reported substantial variability in feature importance across course types, while \citet{veeramachaneniFeatureEngineeringScale2014} and \citet{conijnPredictingStudentPerformance2017} reported wide differences in explained variance across courses, despite using similar behavioural indicators. These findings emphasise portability as a key challenge when applying predictive models beyond their original instructional context.

\subsection{Granularity and alignment with course structure}

In addition to challenges with generalisability, a further limitation of many existing predictive models is their reliance on data from the entire course, which restricts their utility for early intervention. Several studies have explored how accurately student outcomes can be predicted at different points during a course using only the data available up to that week \citep{umerPredictingAcademicPerformance2017, okuboNeuralNetworkApproach2017, taylorLikelyStopPredicting2014}. While they consistently report that accuracy improves as more data accumulates, they demonstrate that early prediction is feasible, with moderate classification accuracy often achieved 3 or 4 weeks into the course \citep{huDevelopingEarlyWarning2014, akcapinarUsingLearningAnalytics2019}. However, these models also reveal necessary trade-offs, as achieving high recall (correctly identifying students who are at risk) often comes at the cost of low precision, meaning that many students flagged for intervention would ultimately succeed without support \citep{conijnPredictingStudentPerformance2017}.

These limitations reflect not only when predictions are made, but also how engagement is measured. Coarse, course-wide indicators lack the temporal resolution needed for weekly predictions. Engagement metrics used for early intervention must be time-sensitive, and capable of capturing how student activity evolves throughout the course. This requires careful attention to granularity. As \citet{sinatraChallengesDefiningMeasuring2015a} note, the appropriate level of granularity depends on which aspect of student engagement is being measured. When tracking behavioural engagement over time, overly coarse metrics may miss important shifts, while overly fine ones risk capturing noise \citep{zhangInvestigatingReliabilityAggregate2024}. The goal is to strike a balance between measures that are fine-grained enough to detect meaningful changes, yet stable enough to support reliable interpretation.

Aligning engagement indicators with the structure of course delivery enhances interpretability. Metrics are most meaningful when measured in line with instructional sequencing, such as at weekly or chapter-based intervals \citep{nguyenUsingTemporalAnalytics2018}. Several studies support this approach. \citet{gohLearningManagementSystem2025} found that consistent engagement across structured course components predicted academic success, while \citet{azconaMicroanalyticsStudentPerformance2015} showed that chapter- or activity-level indicators outperformed coarse aggregates for both prediction and timely intervention. Together, these findings suggest that chapter-level aggregation may provide a practical and pedagogically grounded unit of analysis, supporting early warning while preserving meaningful engagement signals.


\section{Methods}
\label{sec:meth}

\subsection{Moodle data}

To evaluate the approach proposed in this study, we applied our engagement metric retrospectively to three undergraduate statistics courses delivered by the Department of Statistical Science at University College London (UCL) across the 2022--23 and 2023--24 academic years. We refer to these courses as FoundationsCourse, ProgrammingCourse, and StochasticCourse.

All three courses combined in-person lectures with online materials. Moodle primarily serves as a repository for learning resources, including lecture notes, videos, quizzes, and problem sheets. Our analysis focuses on weeks of active instruction spanning weeks 1--11, with week 6 a reading week, excluding university holidays and the pre-exam revision period. Table~\ref{table_studentno} summarises the number of students and grade distributions for each course and academic year. In addition to reporting the mean final grade, we include the number of students scoring below 40 and below 50. A final grade below 40 corresponds to a fail (F grade), while scores below 50 (D and F grades) signal all low-performing students. These thresholds are used throughout the analysis to assess the metric’s potential for identifying students who may benefit from early support.

\begin{table}[hbt]
\centering
\renewcommand{\arraystretch}{0.95}
\caption[Student numbers]{\label{table_studentno} Student numbers, mean final grades, and proportions of lower-performing students per course and year.}
\vspace*{1ex}
\begin{tabular}{r l c c c c}
\toprule
Course & Academic Year & Total & Mean Grade & $<$40 & $<$50 \\[1ex]
\midrule 

FoundationsCourse & 2022--23 & 174 & 62.1 & 12 & 39 \\
                  &          &     &      & \textcolor{gray}{\footnotesize\textit{(6.9\%)}} & \textcolor{gray}{\footnotesize\textit{(22.4\%)}} \\
                  & 2023--24 & 186 & 66.4 & 9 & 34 \\
                  &          &     &      & \textcolor{gray}{\footnotesize\textit{(4.8\%)}} & \textcolor{gray}{\footnotesize\textit{(18.3\%)}} \\

ProgrammingCourse & 2022--23 & 150 & 69.7 & 2 & 7 \\
                  &          &     &      & \textcolor{gray}{\footnotesize\textit{(1.3\%)}} & \textcolor{gray}{\footnotesize\textit{(4.7\%)}} \\
                  & 2023--24 & 174 & 71.9 & 2 & 7 \\
                  &          &     &      & \textcolor{gray}{\footnotesize\textit{(1.2\%)}} & \textcolor{gray}{\footnotesize\textit{(4.2\%)}} \\

StochasticCourse  & 2022--23 & 182 & 65.2 & 9 & 25 \\
                  &          &     &      & \textcolor{gray}{\footnotesize\textit{(5.0\%)}} & \textcolor{gray}{\footnotesize\textit{(13.7\%)}} \\
                  & 2023--24 & 168 & 66.8 & 4 & 21 \\
                  &          &     &      & \textcolor{gray}{\footnotesize\textit{(2.4\%)}} & \textcolor{gray}{\footnotesize\textit{(12.5\%)}} \\ [1ex]

\bottomrule
\end{tabular}
\end{table}

\textbf{FoundationsCourse} is a compulsory first-year course introducing probability and statistical theory. In 2022--23, new chapters were released weekly with clearly labelled resources associated with each chapter, while in 2023--24 the structure was revised to five chapters studied over multiple weeks. Assessment is primarily exam-based (75\%), with a participation grade (15\%) for submitting weekly problem sheets, and coursework (10\%).

\textbf{ProgrammingCourse} is a compulsory first-year course that trains students in practical statistical skills using \texttt{R}. New chapters were released weekly, supported by in-person lab coding exercises. Assessment is weighted towards an online quiz (25\%) in week 11 of the analysed period, and a group project (75\%) scheduled later in the academic year.

\textbf{StochasticCourse} is an optional course for second and third-year students, introducing systems that evolve randomly over time. New content is released weekly, accompanied by a broad range of supporting materials on Moodle, which are clearly labelled by chapter. The course leader and overall course structure remained largely consistent across both academic years, providing a stable basis for comparison. Assessment includes a final exam (75\%), two online quizzes in weeks 5 and 11 (15\% combined), and a participation grade (10\%), which requires students to submit a weekly problem sheet for feedback.

Although demographic information is not available at the individual level in our dataset, broader statistics from the university provide context for the likely composition of the student cohorts \citep{uclStudentStatistics2018}. In 2022--23, 50\% of undergraduate students enrolled in the Department of Statistical Science identified as female, falling slightly to 46\% in 2023--24. A substantial proportion of undergraduate students in the department were classified as overseas: 74\% in 2022--23 and 69\% in 2023--24. Of the total undergraduate overseas student population at UCL, the majority identified as ethnically Chinese (53\% in 2022--23 and 58\% in 2023--24). Although we cannot infer precise demographics for our sample, these figures provide context for the likely composition of students enrolled in the courses included. However, demographic information was not considered throughout the analysis. 

\subsubsection{Data processing}

We extracted the log data from the UCL Moodle platform for the three courses across two academic years. The dataset has six columns that provide details such as the click time and the type of action the user takes. Table \ref{table_datcolumns} describes each column.

\begin{table}[hbt]
\centering
\renewcommand{\arraystretch}{1.3}
\caption[Moodle dataset]{\label{table_datcolumns}
Summary of the Moodle activity log data, highlighting the key columns and the information they contain.}\vspace*{1ex}
\begin{tabular}{l p{9cm}}
\hline
Column name & Column description \\[1ex]
\hline
\texttt{Time} & The data and time (to the nearest minute) \\
\texttt{User} & User ID number for anonymity  \\
\texttt{Event.context} & Title of the link or resource accessed \\
\texttt{Component} & The category of activity or resource accessed \\
\texttt{Event.name} & The purpose of the click \\
\texttt{Description} & Details of the click, including user ID and resource ID \\[1ex]
\hline 
\end{tabular}
\end{table}

VLE activity was restricted to each course’s teaching weeks and labelled by week number (1–11) according to the date of access. Chapter labels were extracted from the \texttt{Event.name} field by identifying numeric patterns (e.g., “Chapter 3 Notes”). Manual overrides were applied where necessary, particularly for tutorials, which often followed a different numbering scheme or covered material from prior weeks, and for resources where numeric identifiers did not correspond to chapters.

Students with a recorded final grade or exam grade of zero were excluded from all analyses. These cases typically indicated exam absences, rather than true academic failure.

\subsubsection{Study sessions}
\label{sessions}

In the learning analytics literature, periods of sustained activity, commonly referred to as study sessions, are often extracted from VLE logs as a foundational step for calculating engagement indicators. Adopting this approach, we identify study sessions as the unit of analysis and use them to calculate the three behaviour indicators: \textit{Frequency}, \textit{Immediacy}, and \textit{Diversity} (see Section~\ref{SEM2}). In the course-wide metric proposed by \citet{chongProxyVariablesOnline2019}, these indicators are aggregated across the entire course. In contrast, our chapter-aligned reconceptualisation attributes each session to a specific chapter, enabling each session to be aligned with a specific point in the course structure.

Following standard practice, we define the boundaries of a study session using periods of inactivity in the VLE log data. Previous research identified 5--30 minute gaps as indicative of a break in engagement \citep{munkImpactDifferentPreProcessing2011, delvalleOnlineLearningLearner2009, ba-omarFrameworkUsingWeb2007}. We adopt a fixed threshold tailored to each course by identifying the 95th percentile of inactivity durations within a two-hour window. This yielded thresholds ranging from 7 to 28 minutes across the datasets. For real-time application to a new course, this threshold can be established by computing the 95th percentile of inactivity durations observed up to that point. In practice, we found that subsequent calculations were not sensitive to the exact threshold selected.

To associate sessions with chapters, we infer the topic based on the names of the resources accessed. When a session includes resources clearly labelled with a chapter identifier, we attribute it to that chapter. If multiple chapter-specific resources are accessed, we treat the appearance of a new chapter as the start of a new session. Sessions that include only general materials (e.g., forums, reference documents) or unlabelled resources are excluded from metric calculations. This results in a set of clearly defined, chapter-specific study sessions, which form the basis of our adaptation of the online behaviour engagement metric by linking engagement directly to the course’s instructional design.

\subsection{Online behavioural student engagement metric}

\subsubsection{Course-wide metric}
\label{CW}

\citet{chongProxyVariablesOnline2019} proposed a metric combining five weighted indicators from student VLE log data to quantify online behavioural engagement over a full course. \textit{Immediacy} measured how quickly a student interacted with the material after it became available, calculated as the number of days between the start of the course and the first study session. \textit{Frequency.1} (which we refer to simply as Frequency) counted the total number of online learning study sessions, while \textit{Frequency.2} (renamed Diversity) reflected the number of distinct learning activities a student engaged with via Moodle across all study sessions. \textit{Recency} recorded how many days there were between the final session and the end of the course, and \textit{Interval} represented the span (in days) between the student's first and last online session.

The course-level online engagement metric, \( Y^{(i)} \),  for student \( i \) is then calculated as a weighted sum:

\begin{equation}
Y^{(i)} = w_I \cdot I^{(i)} + w_{F} \cdot F^{(i)} + w_{D} \cdot D^{(i)} + w_R \cdot R^{(i)} + w_{\text{Int}} \cdot \text{Int}^{(i)} \ ,
\label{eq:WC_score}
\end{equation}
where \(I^{(i)} \), \( F^{(i)} \), \( D^{(i)} \), \( R^{(i)} \), \( \text{Int}^{(i)} \) refer to the min-max scaled indicators for Immediacy, Frequency, Diversity, Recency, and Interval, respectively. The authors assigned equal weights to each indicator, implicitly assuming equal importance.

To assess the relationship between the engagement metric and student performance, the authors conducted a one-way analysis of variance (ANOVA), categorising VLE engagement scores into bands determined by deviations from the mean. The ANOVA revealed statistically significant differences in course achievement across varying levels of VLE engagement for all courses studied ($p < 0.001$ for each course). These results provide empirical support for the validity of the metric as a proxy for online behavioural engagement, suggesting that higher VLE engagement is associated with better academic performance.

However, two key limitations remain. First, the metric is inherently retrospective, requiring data from the entire course to calculate final values, which limits real-time monitoring and timely interventions for at-risk students. Second, it treats the course as homogeneous, disregarding underlying temporal structures commonly found in course designs. In higher education, instructional content is often structured and delivered incrementally, sometimes weekly, sequentially, or by clearly defined chapters, with engagement levels expected to fluctuate accordingly. A single aggregated metric may mask meaningful shifts in engagement that occur across different stages of the course \citep{henrieMeasuringStudentEngagement2015}.

To effectively capture online behavioural engagement, a metric must be sensitive to the dynamic nature of student behaviour. \citet{fredricksSchoolEngagementPotential2004} highlighted the need to understand student engagement not merely as static participation, but as ongoing interactions that evolve throughout a learning experience. In their review of online engagement measurement, \citet{henrieMeasuringStudentEngagement2015} also emphasised the importance of aligning indicators closely with the instructional design and temporal structure of courses. They argued that meaningful engagement metrics should reflect how students interact with content as it is sequentially released, and highlighted the limitations of approaches that aggregate data retrospectively without considering the instructional context.

Building upon these insights, our reconceptualisation of the metric adopts a chapter-level granularity, directly linking engagement measures to specific segments of course content. By structuring the metric in alignment with how material is incrementally presented, we ensure that engagement indicators meaningfully reflect changes in student participation throughout the course. This approach not only aligns with the theoretical foundations described by \citet{fredricksSchoolEngagementPotential2004} but also addresses practical measurement challenges identified by \citet{henrieMeasuringStudentEngagement2015}, facilitating deeper theoretical interpretations and practical interventions, such as early warning systems tailored to specific phases of a course.

\subsubsection{Chapter-aligned reconceptualisation}

To overcome the retrospective design and temporal insensitivity of the course-wide metric, we propose a chapter-aligned reconceptualisation that enables both real-time monitoring and finer-grained analysis of how student engagement unfolds throughout the course.

Our approach calculates online behavioural engagement separately for each chapter, enabling the metric to be computed at any point during the course. For simplicity and alignment with common instructional pacing, we assume engagement is measured at weekly intervals, reflecting the typical sequential release of learning materials over time. This provides a consistent temporal resolution suitable for monitoring trends and triggering timely interventions. However, the method remains flexible and can be applied to any course with some form of structured progression, even if not all resources follow a strict weekly release. At each measurement week $t$, the engagement indicators are computed independently for each available chapter, scaled, and summed. This provides a chapter-level engagement score. Subsequently, these chapter-level scores are combined using a weighted sum to produce a continuous, cumulative measure of online behavioural engagement.

However, adapting the metric to a real-time, cumulative context introduces challenges for certain indicators. Specifically, Recency (the time between the last session and the course end) and Interval (the time span between the first and last session) become highly sensitive to the timing of measurement. For example, if engagement is measured weekly on a Sunday, students who typically engage early in the week (e.g., on Mondays or Tuesdays) may appear to have longer Recency gaps than those who study on Saturdays, even if both demonstrate consistent weekly engagement. Additionally, ongoing interactions with earlier chapters after their primary release period, which may indicate revision rather than active synchronous engagement, can misleadingly inflate these indicators and make them unreliable in a cumulative, real-time metric. Given these considerations, we exclude Recency and Interval and instead focus exclusively on indicators that reliably capture timely chapter-specific engagement: Frequency, Immediacy, and Diversity.

For student $i$ in chapter $k$ at measurement week $t$, we compute three chapter-level indicators of engagement. These are calculated based on student activity observed up to and including week $t$ (except for Immediacy, which remains fixed once the student first engages with the chapter). The indicators are:

\begin{itemize}
  \item \textbf{Immediacy}: The time elapsed (in days) between the release of chapter $k$ and the student’s first study session involving resources labelled for that chapter. This value is fixed once observed and does not change with $t$.
  \item \textbf{Diversity}: The number of distinct learning activities accessed by the student during chapter $k$ study sessions up to and including week $t$.
  \item \textbf{Frequency}: The number of study sessions associated with chapter $k$ completed by the student up to and including week $t$.
\end{itemize}

Each raw indicator value is then min-max scaled relative to the distribution of peer behaviour observed up to and including week $t$, resulting in the scaled values $I_{k,t}^{(i)}$, $D_{k,t}^{(i)}$, and $F_{k,t}^{(i)}$, all ranging from 0 to 1. This time-specific scaling ensures that engagement is interpreted in the context of how other students have engaged with the same chapter up to that point in the course. Details of how these indicators are constructed from the data are given in Section~\ref{SEM2}.

The chapter-level engagement score for student $i$ in chapter $k$ at time $t$ is then defined as:

\begin{equation}
IDF_{k,t}^{(i)} = F_{k,t}^{(i)} + I_{k,t}^{(i)} + D_{k,t}^{(i)} \ .
\label{eq:chapter_score}
\end{equation}
We set $IDF_{k,t}^{(i)} = 0$ for any chapter not yet released by week $t$, or if the student has not engaged with the chapter by that time.

Finally, the overall online behavioural engagement metric for student $i$ at week $t$ is computed as a weighted sum of these chapter-level scores:
\begin{equation}
y_t^{(i)} = \sum_k w_k IDF_{k,t}^{(i)} \ ,
\label{eq:overall_score}
\end{equation}
where $w_k$ is the weight assigned to chapter $k$. These weights allow the engagement metric to reflect differences in the relative importance of chapters within the instructional design. In this study, we set all weights equal to 1, treating each chapter as equally important.

\subsubsection{Constructing chapter-level engagement scores}
\label{SEM2}

At week $t$, for each chapter $k$ and student $i$, we calculate three indicators -- Frequency, Immediacy, and Diversity -- based on all study sessions labelled with chapter $k$ (described in Section~\ref{sessions}) and occurring up to and including week $t$.

\paragraph{Frequency} From the log data, we count the total number of distinct study sessions for chapter $k$ up to and including week $t$, denoted $L_{k,t}^{(i)} \in \mathbb{N}_0$. The raw Frequency indicator is then defined as
\[
\tilde{F}_{k,t}^{(i)} = L_{k,t}^{(i)} \ .
\]

\paragraph{Immediacy} We define the set of session day offsets $\{ \delta_{k,t,l}^{(i)} \mid l = 1, \ldots, L_{k,t}^{(i)} \}$, where each $\delta_{k,t,l}^{(i)} \in \mathbb{N}_0$ records the number of days from the start of term to the $l$-th session for chapter $k$, based on data observed up to and including week $t$. The earliest session for student $i$ is then $\delta_{k,t}^{(i)} = \min \{ \delta_{k,t,l}^{(i)} \mid l = 1, \ldots, L_{k,t}^{(i)} \}
$. We approximate the release point for chapter $k$ as the earliest day on which any student accessed a resource from that chapter 
$\delta_{k,t}^{\star} = \min_{i'} \left( \delta_{k,t}^{(i')} \right)$.
The raw Immediacy indicator is defined as
\[
\tilde{I}_{k,t}^{(i)} = \delta_{k,t}^{\star} - \delta_{k,t}^{(i)} \ .
\]
A value of zero indicates the earliest observed engagement with a chapter, while increasingly negative values reflect longer delays relative to that earliest point. Since exact release dates were not available through the Moodle data, $\delta_{k,t}^{\star}$ serves as a proxy based on first-access patterns. This approximation does not affect the final indicator because all values are min-max scaled, meaning the scores depend only on the relative timing of students’ engagement. Even if the exact release date were known and used instead of $\delta_{k,t}^{\star}$, the scaled values would remain the same.

\paragraph{Diversity} Let $N_t$ be the total number of available activities in Moodle up to week $t$ identified through the \texttt{Event.name} column. For each student $i$, we construct binary activity vectors $\{ \mathbf{A}_{k,t,l}^{(i)} \mid l = 1, \ldots, L_{k,t}^{(i)} \}$, where each $\mathbf{A}_{k,t,l}^{(i)} \in \{0,1\}^{N_t}$ indicates which of the $N_t$ Moodle activities were accessed during session $l$ of chapter $k$. We aggregate these using the element-wise maximum function to obtain a summary vector $
\mathbf{A}_{k,t}^{(i)} = \max \{ \mathbf{A}_{k,t,l}^{(i)} \mid l = 1, \ldots, L_{k,t}^{(i)} \}$, where a value of 1 in $\mathbf{A}_{k,t}^{(i)}$ indicates that student $i$ accessed the corresponding activity in at least one session. We then define Diversity as the number of distinct activity types accessed by student $i$ in chapter $k$ up to week $t$,
\[
\tilde{D}_{k,t}^{(i)} = \lVert \mathbf{A}_{k,t}^{(i)} \rVert_1 \ .
\]

Higher values correspond to greater behavioural engagement for each of these three indicators. Each indicator is min-max scaled across all students for chapter $k$ at week $t$, producing scaled values in $[0, 1]$. For example, the scaled Immediacy indicator is given by:
\[
I_{k,t}^{(i)} = \frac{ \tilde{I}_{k,t}^{(i)} - \min_{i} \tilde{I}_{k,t}^{(i)} }{ \max_{i} \tilde{I}_{k,t}^{(i)} - \min_{i} \tilde{I}_{k,t}^{(i)} } \ .
\]
The same transformation is applied to Frequency and Diversity to obtain $F_{k,t}^{(i)}$ and $D_{k,t}^{(i)}$, respectively.

These scaled indicators are combined to form the chapter-level engagement score $IDF_{k,t}^{(i)}$ as described in Equation~\ref{eq:chapter_score}. The overall online behavioural engagement metric $y_t^{(i)}$ is computed as the weighted sum of these chapter-level scores across all chapters released by week $t$, as defined in Equation~\ref{eq:overall_score}. For our analysis, we set all weights equal to 1. Since each $IDF_{k,t}^{(i)} \in [0, 3]$, the total score $y_t^{(i)} \in [0, 3K_t]$, where $K_t$ is the number of chapters released by week $t$. As the course progresses and new chapters are introduced, the upper bound of the engagement metric increases accordingly.

\subsection{Evaluating the chapter-aligned metric}

\subsubsection{Alignment with the course-wide metric}
\label{methods_RQ1}

To examine whether our chapter-based engagement metric captures the same underlying structure as the course-wide metric proposed by \citet{chongProxyVariablesOnline2019}, we compare the two across time using rank correlation. This comparison examines whether students’ relative engagement levels, as measured by our weekly metric, correlate with their rankings under the retrospective, full-course formulation.

We reconstruct the \citet{chongProxyVariablesOnline2019} engagement metric on our dataset, applying their methodology to combine all five indicators across the whole course duration, without reference to chapter structure. This contrasts with our formulation, which uses only three of these indicators and is calculated cumulatively at the chapter level.

For each student \( i \), we identify all study sessions across the course. Similar to the construction of our chapter-based metric in Section~\ref{SEM2}, let \( L^{(i)} \in \mathbb{N}_0 \) denote the total number of sessions; $\{ \delta_l^{(i)} \mid l = 1, \ldots, L^{(i)} \}$
represent the set of session days, where each \( \delta_l^{(i)} \in \mathbb{N}_0 \) gives the number of days from the start of term to the \( l \)-th session; and $\{ \mathbf{A}_l^{(i)} \mid l = 1, \ldots, L^{(i)} \}, \quad \mathbf{A}_l^{(i)} \in \{0,1\}^N$ be the set of binary activity vectors, where each element indicates whether activity \( j \in \{1, \ldots, N\} \) was accessed in session \( l \).

Using these, we define the five raw (unscaled) engagement indicators as:

\begin{align*}
\tilde{I}^{(i)} &= -\min_l \delta_l^{(i)} \quad &\text{(Immediacy)} \\
\tilde{F}^{(i)} &= L^{(i)} \quad &\text{(Frequency)} \\
\tilde{D}^{(i)} &= \sum_{j=1}^N \left( \max_l \left( A_{l}^{(i)} \right)_j \right) \quad &\text{(Diversity)} \\
\tilde{R}^{(i)} &= \max_l \delta_l^{(i)} \quad &\text{(Recency)} \\
\tilde{\text{Int}}^{(i)} &= \max_l \delta_l^{(i)} - \min_l \delta_l^{(i)}. \quad &\text{(Interval)}
\end{align*}

All five indicators are then min-max scaled across students to give ${I}^{(i)}, \ {F}^{(i)}, \ {D}^{(i)}, \ {R}^{(i)}, \ \text{and} \ {\text{Int}}^{(i)}$ which are combined into the overall engagement score \( Y^{(i)} \), as described in Equation~\ref{eq:WC_score}. Note that Immediacy is negated so that higher values correspond to earlier engagement, and Recency is defined as the last day of engagement rather than its proximity to the end of the course. These adjustments have no impact on the final scaled indicators, as min-max scaling preserves the relative values of students.

For each course and academic year, we compute the Spearman correlation coefficient ($\rho$) between our weekly, chapter-based engagement scores and the course-wide metric. Since our metric is updated cumulatively each week and the course-wide metric is fixed at the course level, this yields a time series of Spearman correlations across weeks 1–11. This allows us to assess how closely the dynamic formulation approximates the retrospective metric over time.

\subsubsection{Predictive validity for academic performance}

To evaluate the predictive validity of our chapter-based engagement metric, we assess how well it correlates with students’ final grades, which serve as a proxy for overall academic achievement within each course. Our primary objective is to determine whether the metric reflects differences in academic performance over time, and whether it does so earlier than the course-wide metric.

We conduct this validation in two complementary stages. First, we calculate the Spearman correlation coefficient ($\rho$) between students’ weekly engagement scores and their final course grades. This is done for each week of the course, allowing us to observe how the association evolves. As a benchmark, we also report the correlation between the course-wide engagement metric (computed retrospectively at the end of the course) and final grade. A higher correlation for the weekly chapter-based metric, especially in earlier weeks, would suggest stronger predictive capacity for real-time monitoring.

Second, we complement these correlation-based findings with a distributional analysis. Each week, we rank students based on their chapter-based engagement scores and group them into five equally sized engagement quintiles: Very Low, Low, Moderate, High, and Very High. We then examine the distribution of final grades within each quintile using boxplots. This approach provides a visual summary of whether students with lower engagement scores tend to achieve lower final grades, and how early this pattern emerges.

This dual approach, which combines correlation with visual distributions, provides both statistical and practical validation of the chapter-based engagement metric as an early indicator of academic risk.

\subsubsection{Practical utility for early identification}

To assess the real-world utility of the chapter-based engagement metric for identifying at-risk students, we treat this as a binary classification problem of predicting whether a student will underperform based on their engagement score. We evaluate its predictive performance using three complementary measures: the area under the receiving operating characteristic (ROC) curve (AUC), recall, and precision. These classification performance indicators enable us to quantify how effectively the lowest engagement quintile identifies students likely to underperform, providing practical insight into how the metric can be utilised in early intervention strategies.

We define at-risk students as those who achieve a final grade below 50\% which encapsulates students with a D or F grade. In addition, we separately track students who score below 40\%, representing those who formally fail the course. While AUC provides an overall summary of the metric’s ability to discriminate between low- and high-performing students, it can be misleading in the presence of imbalanced outcome distributions. This is a relevant concern in our context, where the proportion of low-performing students ranges from 4\% to 22\%. AUC can remain high even if the model performs poorly on the minority class, because a few correct rankings can dominate the score. For this reason, we place greater emphasis on recall and precision, which provide more informative measures of early identification performance in imbalanced settings.

Recall reflects the proportion of all low-performing students who fall within the bottom 20\% of engagement scores. In contrast, precision demonstrates the proportion of students in that bottom quintile who ultimately underperform. These measures are calculated on a weekly basis for each course and academic year.

These classification performance indicators are intended to reflect the range of possible outcomes when applying the metric in practice, illustrating the trade-off between early identification and the risk of false positives. They offer a demonstration of how the chapter-based metric could support timely educational interventions, recognising that practical use would also involve contextual information and professional judgment.


\section{Results}
\label{sec:results}

\subsection{Alignment with the course-wide metric}

\begin{figure}[ht]
  \centering
  \includegraphics[width=\textwidth]{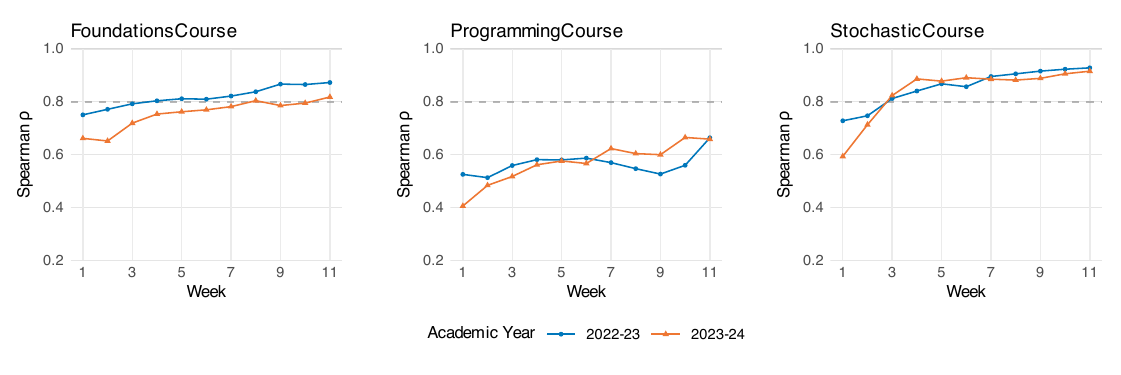}
  \caption{
Spearman correlation ($\rho$) between the weekly chapter-based engagement metric and the course-wide metric, across three courses and two academic years. Blue circles indicate the 2022–23 academic year, orange triangles indicate 2023–24. The dashed grey line marks the $\rho = 0.8$ threshold for strong correlation.
}
  \label{fig:correlation_wc}
\end{figure}

Figure~\ref{fig:correlation_wc} presents the Spearman correlation coefficients comparing weekly chapter-based engagement metrics to the course-wide metric, tracked across weeks and separated by academic year and course. As expected, correlations increase steadily across all course-year combinations as the chapter-based metric accumulates more data. By the final weeks of the term, both years of the FoundationsCourse and StochasticCourse exceed $\rho = 0.8$, a commonly used benchmark for strong monotonic association, indicating strong agreement between chapter-level and course-level engagement rankings.

Early and sustained alignment is most evident in the StochasticCourse, where correlations surpass $\rho = 0.8$ by week 3 and reach $\rho = 0.93$ (2022–23) and $\rho = 0.92$ (2023–24) by week 11. The FoundationsCourse also shows good agreement. In 2022–23 it exceeds $\rho = 0.8$ by week 4 and reaching $\rho = 0.87$ by the end of the term, while the 2023–24 cohort, though consistently slightly lower, reaches $\rho = 0.8$ by week 8.

The ProgrammingCourse shows lower correlations throughout, ranging from $\rho = 0.54$ to 0.75 in 2022–23 and from $\rho = 0.40$ to 0.67 in 2023–24. Although these correlations increase over time, they plateau below the levels observed in other courses. Section~\ref{sec:discussion} explores possible explanations for this discrepancy.

These results demonstrate that the chapter-based metric can reliably approximate the full-course engagement measure early in the term, capturing students’ relative engagement by week 3 in some cases, although this varies by course.

\subsection{Predictive validity for academic performance}

\subsubsection{Correlation with final grade}
\begin{figure}[ht]
  \centering
  \includegraphics[width=\textwidth]{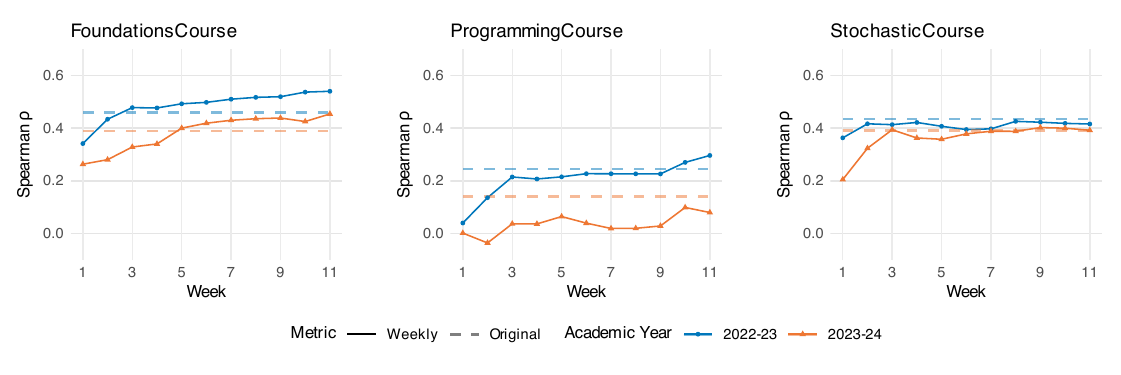}
  \caption{
Spearman correlation ($\rho$) between engagement metrics and final grades. Each of the three courses includes two academic years: 2022–23 is represented by blue circles, and 2023–24 by orange triangles. Two engagement metrics are displayed: the chapter-based metric (solid lines), calculated weekly, and the course-wide metric (dashed lines), calculated at the end of the term, which serves as a constant reference.
}
  \label{fig:grade_correlation}
\end{figure}

Figure~\ref{fig:grade_correlation} shows how strongly the weekly chapter-based engagement metric correlates with students' final grades, compared to the course-wide metric. Across most course-year combinations, the weekly metric matches or outperforms the course-wide metric, sometimes early in the term.

The FoundationsCourse exhibited the best predictive performance. In 2022–23, the weekly metric surpassed the course-wide by week 3 and continued to improve, reaching a peak correlation of $\rho = 0.54$ (compared to the course-wide metric's $\rho = 0.46$). In 2023–24, it reached the course-wide metric at week 5 ($\rho = 0.39$) and exceeded $\rho = 0.45$ by the end of the term.

The StochasticCourse showed consistent performance over both academic years. In 2022–23, correlations stabilised between $\rho = 0.40$ and $0.43$ from week 2 onwards, aligning closely with the course-wide metric ($\rho = 0.44$). Similarly, in 2023–24, the correlations ranged from $\rho = 0.36$ to $0.40$ from week 3, remaining comparable to the course-wide metric ($\rho = 0.39$).

The ProgrammingCourse exhibited consistently weak correlations between the engagement metrics and final grades across both years. In 2022–2023, the weekly metric stabilised around $\rho = 0.23$, reaching a maximum of $\rho = 0.30$ at the end of the term, marginally outperforming the course-wide metric ($\rho = 0.25$). In 2023–2024, the weekly metric remained persistently low, never exceeding $\rho = 0.10$, while the course-wide metric was only $\rho = 0.14$. These findings indicate that neither the weekly nor the course-wide engagement metric is a reliable predictor of final grades in the ProgrammingCourse.

These results indicate that the chapter-based metric offers a timely proxy for online behavioural student engagement that is predictive of academic performance in structured, labelled courses. However, its effectiveness appears more limited in the ProgrammingCourse, where the alignment between online engagement and assessed performance may differ due to the skill-based nature of the content and the absence of an invigilated final exam. In such contexts, behavioural engagement with materials may not map as directly onto summative performance as in more traditional, lecture-based statistics courses.

\subsubsection{Grade distributions by engagement quintile}
Using boxplots to further investigate the relationship between engagement and academic performance, we examined the distribution of final grades across engagement quintiles. Although this analysis was conducted for all weeks, we highlight weeks 3 and 6 as illustrative time points. Week 3 is the earliest point where the chapter-based engagement metric begins to approximate the predictive strength of the course-wide metric, while week 6, at the end of the reading week, represents a practical midpoint for educational interventions. Although we focus here on weeks 3 and 6, similar patterns are observed across all weeks.

\begin{figure}[ht]
\centering
\includegraphics[width=\textwidth]{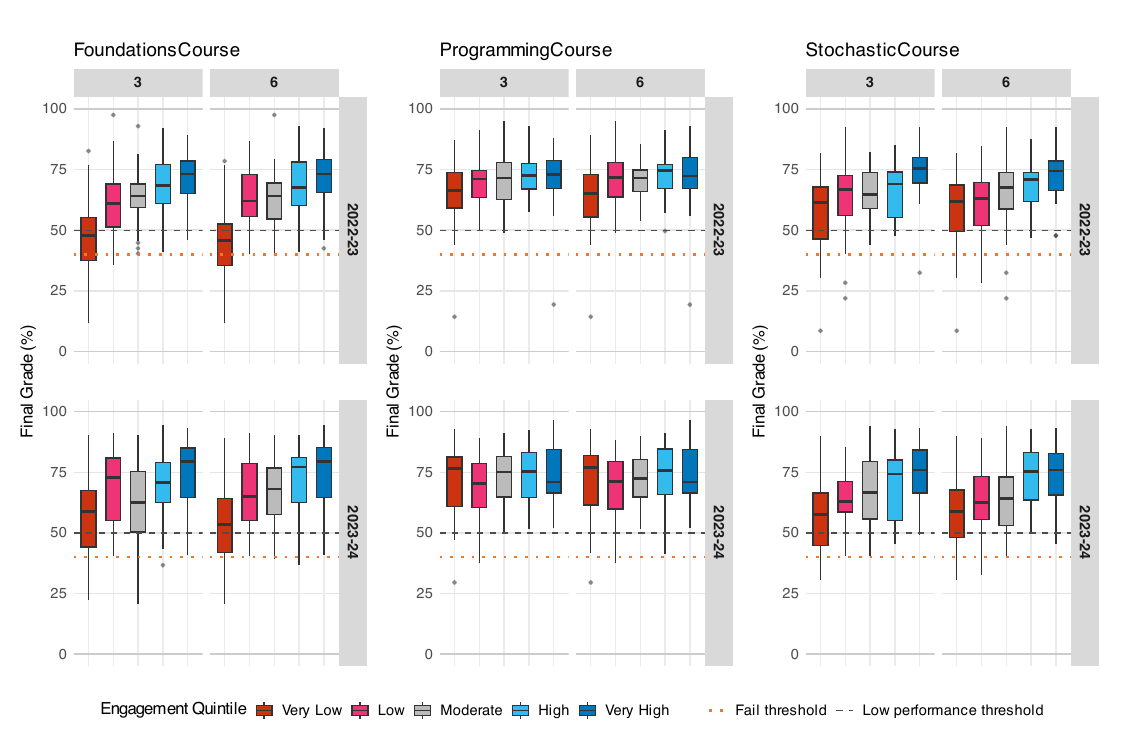}
\caption{Distributions of final grades for each course across two academic years, grouped by student engagement quintiles at weeks 3 and 6. Boxplots display final grade distributions within quintiles ranging from Very Low to Very High based on the weekly chapter-based engagement metric. The dashed black line represents the low-performance threshold (50\%), while the dotted orange line indicates the failure threshold (40\%).}
\label{fig:grade_boxplot}
\end{figure}

Each week, students are categorised into five evenly sized engagement quintiles determined by their weekly chapter-based metric: Very Low, Low, Moderate, High, and Very High. These quintiles allow for a relative comparison of engagement within each cohort. The resulting boxplots (Figure~\ref{fig:grade_boxplot}) display the distribution of final grades within each engagement quintile, with horizontal reference lines at the 50\% (low performance) and 40\% (fail) thresholds.

For the ProgrammingCourse, the boxplots reveal similar grade distributions across all quintiles, with little variation in medians at either week 3 or week 6, for both academic years. This suggests no meaningful association between observed online behavioural engagement and final grades in this course. Furthermore, the small number of students below performance thresholds (only two failing and seven low-performing students per year) limits the effectiveness of the metric in identifying at-risk individuals. As a result, we exclude the ProgrammingCourse from further analysis and revisit potential explanations in Section~\ref{sec:discussion}.

In contrast, the FoundationsCourse and StochasticCourse show a clearer relationship between engagement and academic performance. The median final grade increases with each successive quintile in most cases, although some overlap is observed between adjacent middle quintiles. The Very High engagement group consistently had the highest median grades (ranging from 73.2\% to 79.5\%), and the minimum interquartile range observed was 62.5\%, well above the low-performance threshold.

Conversely, the Very Low engagement group consistently had the lowest median grades (ranging from 45.8\% to 61.9\%). Notably, in the 2022–23 FoundationsCourse, this group's median grade remained below the 50\% threshold at both weeks 3 and 6, and included all students who ultimately failed ($<40\%$). Across both the FoundationsCourse and StochasticCourse in both academic years, the Very Low group's interquartile range was the only one to span the 50\% line, further highlighting its ability to flag low performance even at early time points.

Together, these patterns suggest that students with higher online engagement, captured by our chapter-based metric, tend to achieve better academic outcomes. Meanwhile, students in the Very Low group can often be identified as early as week 3 and are more likely to fall below performance thresholds. While the separation among intermediate engagement groups is less consistent, the distributional patterns nonetheless suggest that interventions targeted at the lowest quintile could have the greatest potential impact.

\subsection{Practical utility for early identification}

\begin{figure}[ht]
  \centering
  \includegraphics[width=0.66\textwidth]{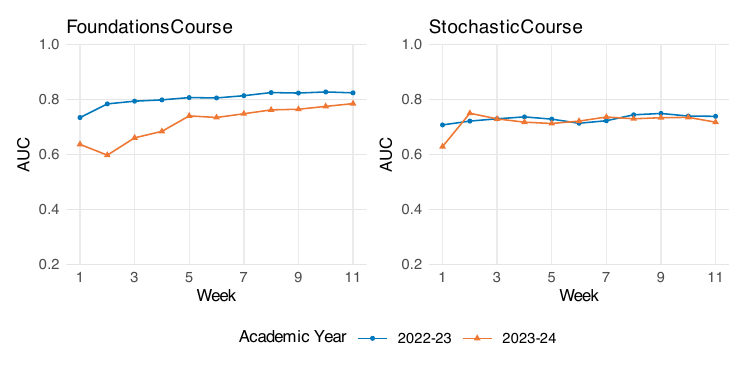}
  \caption{
Area under the ROC curve (AUC) for predicting low performing students ($<50\%$ final grade) using the weekly chapter-based engagement metric. Lines represent AUC values calculated weekly, separated by academic year (blue circles = 2022–23, orange triangles = 2023–24). Higher AUC values indicate stronger discriminatory power of the engagement metric for identifying at-risk students; an AUC of 1.0 reflects perfect prediction, while an AUC of 0.5 indicates performance no better than random chance. Results are shown for the FoundationsCourse and StochasticCourse; the ProgrammingCourse is excluded due to a lack of meaningful association between engagement and final grade.}
  \label{fig:auc}
\end{figure}

\begin{figure}[ht]
  \centering
  \includegraphics[width=0.66\textwidth]{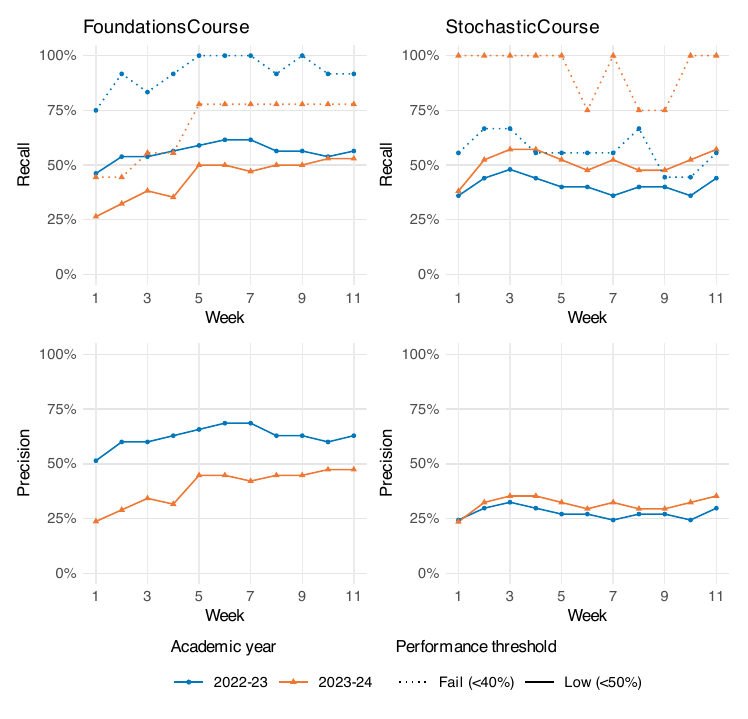}
  \caption{
Recall and precision for identifying low performing students (final grade $<50\%$) based on the lowest engagement quintile, plotted across university weeks for each course and academic year (2022–23 = blue circles, 2023–24 = orange triangles). Each column corresponds to a different course. The top row shows recall, which is the proportion of all low-performing students (final grade $<50\%$) captured within the bottom 20\% of engagement scores. An additional dotted line indicates recall for failing students (final grade $<40\%$). The bottom row shows precision, which is the proportion of students in the bottom 20\% who ultimately scored below 50\%. Higher values indicate more effective early identification of at-risk students using the chapter-based engagement metric.}
  \label{fig:recall_precision}
\end{figure}

We evaluate the practical usefulness of the bottom quintile for early identification of at-risk students by examining its predictive performance using AUC (Figure~\ref{fig:auc}), recall, and precision (Figure~\ref{fig:recall_precision}). 

The FoundationsCourse demonstrated the most impressive AUC in 2022–23, surpassing 0.8 by week 4 and steadily rising to 0.825. The subsequent year, 2023–24, was generally lower, but gained consistency after week 4, where it ranged between 0.672 and 0.771. The StochasticCourse displayed stable year-to-year performance, with AUC values consistently between 0.718 and 0.750 from week 3 onward. However, it must be noted that these values can be unreliable when dealing with unbalanced groups.

Recall refers to the proportion of students who scored below 50\% in the final grade and were correctly identified as belonging to the Very Low engagement quintile. In 2023–24, both the FoundationsCourse and StochasticCourse hover around the 50\% mark for identifying low-performing students after the midpoint of the term, with recall rates between 75\% and 100\% for those who go on to fail. In 2022–23, the FoundationsCourse demonstrated generally higher recall, reaching 62\% for low-performing students by week 7 and often identifying 100\% of those who failed. In contrast, the StochasticCourse exhibited lower recall overall, with a minimum of 36\% for low-performing students and around 50\% for those who failed. Although there is some variation across courses and years, the chapter-based metric consistently identifies a meaningful proportion of low-performing students, especially those at risk of failing.

Precision refers to the proportion of students in the Very Low quintile who ultimately scored below 50\%. The results indicate that the metric is most accurate in the FoundationsCourse 2022–23 cohort, where over 60\% of students flagged as low-engaging subsequently had low performance, peaking at 69\% during weeks 6 and 7. In 2023–24, precision in the same course decreased but exceeded 30\% from week 3, reaching a peak of 45\% in the latter half of the term. In contrast, the StochasticCourse demonstrated lower precision overall, ranging between 24\% and 35\% across both years. This disparity reflects the smaller proportion of low-performing students in the StochasticCourse compared to the FoundationsCourse, meaning that although recall is similar, a larger share of flagged students in the StochasticCourse do not fall below the 50\% final grade threshold.

To illustrate the real-world implications of applying this metric for early intervention, we simulate two contrasting scenarios: a best-case and a worst-case scenario. The best case arises from the FoundationsCourse 2022–23, where an intervention at week 6 targeting the bottom 20\% of engagers (35 students) successfully identifies all 12 failing students along with 12 additional low performers, while missing 15 other low-performing students and incorrectly flagging 11 students who ultimately performed adequately. In contrast, a worst-case scenario is observed in StochasticCourse 2022–23, where intervening at week 7 identifies only 5 of the 7 failing students and 4 additional low performers, while missing 16 low-performing students (including 2 failures) and unnecessarily flagging 28 students. These examples illustrate the trade-offs between recall and precision in practice, underscoring that while the chapter-based metric can be highly effective in specific settings, its performance depends on the course and year.

These findings highlight both the challenges and opportunities of early identification. This real-time, chapter-based metric, drawn from Moodle logs without modelling, can identify, on average, half of the students at risk of low performance from the midpoint of the course. While precision remains moderate, particularly in cohorts with a small number of low-performing students, the high recall suggests that this approach could guide initial support efforts.


\section{Discussion and Conclusion}
\label{sec:discussion}

In this paper, we reformulated the course-wide engagement metric proposed by \citet{chongProxyVariablesOnline2019} into a dynamic, chapter-based version that can be computed iteratively throughout a course. We evaluated this revised metric by examining whether it captures similar behavioural signals to the course-wide formulation, how early in the course it begins to predict students’ final grades, and how it can be used to support early identification and intervention for at-risk students.

In the FoundationsCourse and StochasticCourse, strong alignment between the weekly and course-wide metrics was observed from as early as week 3. This suggests that excluding the course-wide metric’s Recency and Interval indicators and integrating the chapter structure did not compromise its capacity to detect meaningful engagement patterns. Moreover, correlations with final grades surpassed or matched the course-wide version by week 5, indicating the revised metric demonstrates predictive utility from halfway through the term.

The bottom engagement quintile typically included around half of students who scored below 50\%, and an even greater share of those who failed (below 40\%). However, many students flagged as low-engaging ultimately passed, highlighting the limited precision of this approach. This mirrors a common challenge in early-warning systems in balancing sensitivity (detecting at-risk students) with specificity (avoiding false positives) \citep{conijnPredictingStudentPerformance2017, umerPredictingAcademicPerformance2017}.

Performance varied by course and year. The StochasticCourse demonstrated consistent results across both years, with stable alignment between metrics, grade correlations, and recall-precision trade-offs. This stability may be due to its tightly controlled instructional design, which remained unchanged between cohorts. Prior research has emphasised the challenge of cross-cohort generalisability due to variation in platforms, content, or teaching styles \citep{boyerTransferLearningPredictive2015, dingTransferLearningUsing2019, motzValidityUtilityActivity2019}. However, our findings suggest that stable course delivery can support comparability in engagement metrics over time.

The FoundationsCourse showed greater variation between the academic years. In 2022–23, it outperformed the following year throughout the analysis. A likely reason is the structural reorganisation of the course in 2023–24, where chapters were merged and delivered across multiple weeks, weakening alignment between behavioural data and instructional units. Prior studies have stressed that engagement metrics are most effective when aligned with pedagogical structure \citep{gasevicLearningAnalyticsShould2016, nguyenUsingTemporalAnalytics2018}. Our findings reinforce that the metric performs best when engagement is measured at a level of granularity that reflects the course’s intended sequencing.

In contrast to both the StochasticCourse and the FoundationsCourse, the ProgrammingCourse showed consistently weaker results. Correlations between the weekly and course-wide metrics remained below 0.7 throughout the term, including in the final week when both metrics drew on the same underlying data. Further analysis revealed that removing the Recency and Interval indicators from the course-wide metric substantially increased the alignment between metrics ($\rho > 0.8$ by week 8), suggesting that these indicators drove the discrepancy. In this course, an assessment was conducted during the final week, and most students (100\% in 2022-23 and 97\% in 2023-24) interacted in the VLE that week. However, the Recency values varied considerably depending on the exact day of interaction, despite this variation holding little practical significance since the assessment effectively marked the end of the course. As such, these indicators introduced noise without capturing meaningful behavioural differences.

Regardless of these differences, both metrics’ predictive performance was poor, and the chapter-based engagement quintiles failed to distinguish between grade distributions. This breakdown likely stems from a combination of factors, though we cannot isolate their relative influence without a broader sample.

One possible explanation relates to assessment design. Students were graded on a take-home multiple-choice quiz and a group project, both of which undermine the validity of student grades as a proxy for academic performance. Group assessments can obscure individual ability 
\cite{almondGroupAssessmentComparing2009}, and non-invigilated online quizzes raise concerns regarding academic integrity. These limitations may have been further exacerbated in 2023–24 with the increased accessibility of generative AI tools such as ChatGPT.

A second consideration is the nature of the course activity. The course demanded extensive work in R, a programming environment external to the VLE, where interactions were not captured in the log data. Consequently, many sessions likely involved behavioural engagement, but were not fully captured because they occurred outside the VLE. This reflects a broader limitation of VLE-based engagement metrics as they can only measure observable interactions within the VLE platform and may miss substantial off-platform activity \citep{martinOnlineLearnerEngagement2022}.

Finally, the relationship between behavioural engagement and academic achievement may differ in skill-based or technical courses. Students began the ProgrammingCourse with varying levels of prior programming experience. Less experienced students likely needed to engage more frequently with course materials to meet expectations, whereas those with more experience could interact minimally and still achieve high grades. In such contexts, high behavioural engagement may signal struggle rather than mastery \citep{beckWheelSpinningStudents2013}, which would lead to misclassification in early-warning systems based on behavioural indicators.

\subsection{Practical Implications}

The engagement metric proposed in this study offers a practical and interpretable framework for monitoring student activity in real-time, based on the chapter-based structure of online courses. Unlike predictive models that require training data and modelling, this metric can be utilised from the outset of a course and across various educational settings with minimal technical infrastructure. Its transparency and adaptability make it particularly valuable for institutions seeking scalable learning analytics solutions without significant investment in modelling pipelines.

A central advantage of the metric lies in its generalisability. While predictive models typically require training on historical data and often fail to generalise across academic years or course contexts \citep{boyerTransferLearningPredictive2015, dingTransferLearningUsing2019}, this metric can be applied consistently from one cohort to the next. Our findings suggest that when instructional design and delivery remain stable, the performance of the metric is also relatively consistent, particularly in more traditional, lecture-based courses. This also means that historical data from previous years can serve a valuable role by indicating when the metric tends to stabilise and how it correlates with student outcomes, providing a practical reference for future cohorts.

We demonstrated a straightforward way to incorporate the metric into an early-warning strategy by identifying students in the bottom 20\% of cumulative engagement at various points in the course. Although the method is relatively coarse, it effectively identified a substantial proportion of students who ultimately scored below 50\%, including most of those who failed. However, it had low precision, leading to many false positives. Whether this trade-off is acceptable depends on the type of intervention. For example, a broader identification approach might be suitable for low-resource strategies such as automated messaging or behavioural nudges.

Beyond its utility as an aggregate score, the metric is also interpretable through its construction, providing more granular diagnostic insights. Engagement can be tracked weekly, chapter by chapter, or decomposed into indicators that reflect the Immediacy, Frequency, and Diversity of interactions. These dimensions can reveal engagement trajectories and behavioural shifts, such as falling behind the intended pace or interacting with only a narrow range of materials, providing instructors or students with timely, actionable information. Future work should explore how these indicators can be made accessible through dashboards or feedback systems for targeted support and informed pedagogical adjustments.

\subsection{Limitations}

This study evaluated the metric using only three undergraduate modules over two academic years. While this provided initial insight into variations across cohorts and course types, the limited scope constrains transferability. We observed performance differences between courses and years; however, given the small sample size, we cannot draw firm conclusions about the underlying causes. Although our results support hypotheses concerning course structure, assessment design, and instructional alignment, these must be tested across a broader and more diverse evidence base, as differences in performance between modules may reflect course-specific characteristics rather than broader patterns. Without a more diverse sample that spans disciplines, delivery formats, and institutional contexts, conclusions about the metric’s validity remain provisional.

Secondly, the analysis focused solely on online behavioural engagement, the dimension most accessible through VLE log data. This excludes other forms of engagement, such as emotional and cognitive, which are known to influence academic outcomes \citep{fredricksSchoolEngagementPotential2004, fredricksStudentEngagementContext2016, caspari-sadeghiApplyingLearningAnalytics2022}. Consequently, the metric may overlook disengaged students whose behavioural patterns remain superficially active. Future work should investigate data triangulation by integrating behavioural data with self-report or textual analysis data to develop more reliable and encompassing engagement metrics \citep{okurBehavioralEngagementDetection2017, martinOnlineLearnerEngagement2022}.

In addition, this study evaluated the metric solely in relation to final grades, treating academic performance as the primary outcome. However, student engagement is linked to a broader range of outcomes, including motivation, self-regulation, and satisfaction \citep{taiBewareSimpleImpact2019, luOnlineLearningMeanings2020}. As such, our findings may underestimate the value of identifying students with low behavioural engagement. Future work should explore how the metric relates to these additional dimensions of the student experience. Identifying and supporting disengaged students may prove beneficial even when their academic performance is not immediately at risk.

Finally, the utility of the metric relies heavily on well-defined course sequencing. Engagement scores are sensitive to how VLE resources relate to chapters. Without consistent naming conventions, misclassification can compromise the validity and usefulness of the metric. Therefore, applying the metric requires either structured resource labelling or extensive manual intervention, both of which pose scalability challenges. Institutions adopting chapter-based analytics should prioritise alignment between pedagogical design and VLE architecture.

Overall, these findings demonstrate that chapter-based engagement metrics provide a scalable, interpretable, and low-resource alternative to retrospective or model-based approaches for monitoring online behavioural student engagement. When course design aligns closely with VLE structure, the metric proved effective at capturing early patterns of engagement and identifying students at risk. This reinforces the potential of chapter-based metrics as practical tools for early-warning systems, particularly in structured, lecture-based modules. More broadly, the findings emphasise the importance of integrating engagement measurement into pedagogical design, rather than treating it as a post-hoc analytical layer. As institutions continue to invest in real-time learning analytics, future work should validate these findings across a wider range of course contexts and explore how such metrics can be embedded into feedback systems, dashboards, and targeted interventions to support both students and instructors throughout the learning process.


\bibliography{SEM}

\end{document}